\documentstyle[epsf,aps,prl]{revtex}

\begin{document}
\draft

\twocolumn[
\hsize\textwidth\columnwidth\hsize\csname@twocolumnfalse\endcsname

\title{Metastability and the Casimir Effect in Micromechanical Systems } 

\author{E. Buks and M. L. Roukes}

\address{Condensed Matter Physics, California Institute of Technology, Pasadena, CA 91125} 

\date{\today} 
 
\maketitle 
 
\begin{abstract} 
Electrostatic and Casimir interactions limit the range of positional stability of 
electrostatically-actuated or capacitively-coupled mechanical devices.  We investigate 
this range experimentally for a generic system consisting of a doubly-clamped Au 
suspended beam, capacitively-coupled to an adjacent stationary electrode. The mechanical 
properties of the beam, both in the linear and nonlinear regimes, are monitored as the 
attractive forces are increased to the point of instability. There "pull-in" occurs, 
resulting in permanent adhesion between the electrodes. We investigate, experimentally 
and theoretically, the position-dependent lifetimes of the free state (existing prior 
to pull-in). We find that the data cannot be accounted for by simple theory; the 
discrepancy may be reflective of internal structural instabilities within the 
metal electrodes. 
\end{abstract} 


] The technology of micro electro mechanical systems (MEMS) now routinely
allows fabrication of miniature movable structures {\it on-chip}. This new
branch of microelectronics enables the realization of a variety of
miniature, fully integrated sensors and actuators, with a rapidly growing
range of applications. \ One common building block within MEMS is the
capacitively-tunable resonator (CTR), comprising a movable electrode that
serves as a mechanical resonator, and an adjacent stationary electrode (see 
\cite{15} and references therein). \ The capacitive coupling between the
electrodes allows electrostatic control of both the mean position of the
resonator and its resonance frequency. \ In addition to the electrostatic
interaction, such electrodes are coupled via the Casimir effect \cite{793}, 
\cite{850}. \ This quantum electrodynamical effect originates from the
dependence of the ground state energy of the electromagnetic field upon
boundary conditions and leads to an observable forces between macroscopic
bodies. \ Due to its relatively short range, this force has only a small
effect on the dynamics of macroscopic mechanical systems. \ However, the
Casimir force can play a major role in modern MEMS where typical distances
between neighboring surfaces can be on the sub-micron length scale \cite
{2501}. \ These attractive electrostatic and Casimir forces give rise to a
mechanical metastability that may cause {\it stiction} (for a review see 
\cite{385}, \cite{1}). \ In the present case it is manifested as collapse (%
{\it pull-in}) of the movable resonator onto the nearby electrode, resulting
in their permanent adhesion. \ This phenomena can be a principal cause of
malfunctioning in MEMS. \ Moreover, it limits the range of stable operation
of CTRs.

In the present work we study experimentally and theoretically the range of
tunability and mechanical metastability in a CTR system made\ of a doubly
clamped Au beam separated from an adjacent counterelectrode by a small
vacuum gap. \ Once the beam is brought to contact with the electrode,
permanent adhesion occurs. \ This indicates that the {\it free} state of
this system is merely metastable, and that the state of {\it contact} (after
pull-in) has lower energy due to the strongly attractive Casimir
interaction. \ The potential barrier separating these two states determines
the lifetime of the free state. \ The barrier can be reduced by introducing
electrostatic attraction, namely by applying a DC voltage, $V_{dc}$%
\thinspace\ between the beam and the electrode. \ In our measurements we
study the mechanical properties of the beam as $V_{dc}$ is gradually
increased to the point of pull-in. \ Near this critical point all of the
measured properties of the beam show strong dependence on $V_{dc}$. \ Using
a simple model we calculate the shape of the potential barrier confining the
metastable mechanical state, and then estimate the escape rate via both
thermal excitation and quantum tunneling.

Fig. 1(a) is a micrograph displaying a side view of our device. \ The
details of the fabrication process are given elsewhere \cite{paper1}. \ The
structure is designed to allow full characterization of the beam's
properties to yield straightforward and unambiguous interpretation of our
results. \ We use bulk (rather than surface) micromachining, which allows
the substrate to be completely removed beneath the sample. \ This greatly
simplifies the boundary conditions of the electromagnetic field in the
vicinity of the sample. \ Moreover, we avoid using multilayered structures,
since their internal stresses are generally important,  and are difficult to
model theoretically. \ The beam has length $l=200$ $\mu $m, width $a=0.28\mu 
$m and thickness $t=0.25\mu $m (measured using an atomic force microscope).
\ At its center, the stationary electrode has a 20$\mu $m long rectangular
protrusion. This is separated from the beam by a vacuum gap of  $g=1.28$ $%
\mu $m.

All measurements are done at room temperature, {\it in-situ} within a
commercial scanning electron microscope (SEM). \ A voltage $%
V=V_{dc}+V_{ac}\cos \left( 2\pi \nu t\right) $ is applied between the beam
and the electrode (see Fig. 1(b)). \ The static response of the beam is
measured by operating the SEM in scanning mode, imaging the device, and
using digital image processing to extract experimental data.\ \ The
dynamical response is measured with a stationary electron beam focused at a
point near the edge of the Au beam. \ To detect mechanical displacement we
use a spectrum analyzer to monitor the output signal from the
photomultiplier serving as a secondary electron detector. \ Note that this
detection scheme is sensitive almost exclusively to motion in the plane of
the sample.

To characterize the mechanical properties of the beam, namely stress and
stiffness, we measure its resonance frequencies. \ For small stiffness they
are given by:

\begin{equation}
\nu _{n}=n\nu _{0}\left[ 1+2\zeta +\left( 4+n^{2}\pi ^{2}/2\right) \zeta ^{2}%
\right] ,  \label{fre}
\end{equation}
where \smallskip $\zeta ^{2}=Ea^{3}t/12Tl^{2},$ $\nu _{0}=\sqrt{T/\rho A}/2l$%
, with $E$ being Young's modulus, $T$ is the stress, and $\rho $ is the mass
density \cite{paper1},\cite{elas}. \ The dimensionless parameter $\zeta $
indicates the relative effect of stiffness compared to stress on the
dynamics of the beam. \ Note that the terms that make the spectrum unequally
spaced are of order $O\left( \zeta ^{2}\right) .$

We excite the beam by applying AC voltage ($V_{dc}=0)$ and find three modes
with frequencies $\nu _{1}=185.53%
\mathop{\rm kHz}%
$, $\nu _{2}=372.4%
\mathop{\rm kHz}%
$, and $\nu _{3}=563.8%
\mathop{\rm kHz}%
$. \ Using Eq. (\ref{fre}) we find $\zeta $=0.017$\pm 0.002$, thus stiffness
has only a small effect on the dynamics of the beam. \ In what follows, we
ignore altogether corrections due to stiffness. \ The stress is found from
the measured resonance frequencies to be $T=7.4\times 10^{-6}%
\mathop{\rm N}%
.$

Next we gradually increase $V_{dc}$ and measure the mechanical properties of
the beam until pull-in occurs. \ Figure 2(a) shows the normalized static
deflection of the beam's center, $x_{1}/g$, as a function of $V_{dc}$. \
After completing the measurements at $V_{dc}=29$ V, pull-in occurred with
this particular beam. \ Simultaneously applying the small AC excitation with
the DC bias allows us to measure the frequency of the fundamental mode $\nu
_{1}=\omega _{1}/2\pi $. \ Figure. 2(b) shows the ratio $-\Delta \omega
/\omega _{0}=-\left( \omega _{1}-\omega _{0}\right) /\omega _{0}$, where $%
\omega _{0}$ is the angular frequency at $V_{dc}=0$. \ Both $x_{1}$ and $%
\omega _{1}$ show strong dependence on $V_{dc}$ near pull-in. \ 

Further insight can be gained by studying the dynamics of the beam in the
nonlinear regime. \ For oscillations with a large amplitude the harmonic
approximation breaks down and effects due to higher order terms of the
potential are observable. \ Figure 3 shows the measured response of a beam
as a function of frequency around the fundamental resonance for a discrete
range of ac drive ($V_{ac})$ levels, at fixed DC bias, $V_{dc}=20$ V. \ As $%
V_{ac}$ is increased, the response becomes asymmetric, with a sharp drop
occuring on the high frequency side. \ Moreover, a strong hysteresis is
found for high $V_{ac}$ when the frequency is scanned back and forth across
the resonance (see inset of Fig. 3). \ Theoretically, the response function
in the nonlinear regime is given to lowest order by \cite{mech}:

\begin{mathletters}
\begin{equation}
a^{2}\left[ \left( \omega -\omega _{0}\left( 1+\kappa a^{2}\right) \right)
^{2}+\left( \omega _{0}/2Q\right) ^{2}\right] =\left( \frac{f}{2m\omega _{0}}%
\right) ^{2},  \eqnum{2}  \label{response}
\end{equation}
where $f$ is the amplitude of the external force, $a$ is the amplitude of
oscillations, $m$ is the mass, and $\kappa $ is a parameter describing the
amplitude-dependent frequency shift due to nonlinearity.

From the shape of the resonance peaks in the nonlinear regime we extract the
nonlinear frequency-shift parameter, $\kappa $, as a function of $V_{dc}$
(Fig. 2(c)). \ For relatively small $V_{dc}$ we find $\kappa >0$. \ For
higher $V_{dc}$ $\kappa $ becomes negative, and close to the point of
pull-in $\kappa $ drops quite rapidly as a function of $V_{dc}$, indicating
the strong nonlinearity of the potential in this regime.

We describe our system using a simple one dimensional model as proposed in
Ref. \cite{193} (see Fig. 4(a)). \ This model employs rather simplistic
approximations and thus cannot be expected to provide accurate quantitative
predictions -- it is extremely useful, however, in providing physical
insight and a qualitative description of the dynamics. \ A mass $m$ coupled
to a string with a constant $k=m\omega _{0}^{2}$ represents the
doubly-clamped beam. \ An adjacent surface at distance $g$ interacts with
the mass electrostatically, as well as by the Casimir effect. \ For our
device geometry it is appropriate to model these forces, $F$, as applied
locally to the beam's center. \ In this case the displacement of the beam's
center is $Fl/4T$, \ and the effective spring constant is thus $k=4T/l$.

The total potential energy is given by:

\end{mathletters}
\begin{equation}
U_{t}\left( x\right) =U_{E}\left( x\right) +U_{V}\left( g-x\right)
+U_{C}\left( g-x\right) .  \label{u1}
\end{equation}
The first term $U_{E}\left( s\right) =\left( 1/2\right) ks^{2}$ is the
beam's elastic potential. \ The second term is the electrostatic potential
introduced by applying a DC voltage, $V$, between the beam and its
counterelectrode. \ We estimate the capacitance $C$ between the beam and the
electrode using a simple parallel plate approximation, thus $U_{V}\left(
s\right) =-\epsilon _{0}AV^{2}/2s$, where $\epsilon _{0}$ is the vacuum
permittivity and $A$ is the area. \ The third term is the Casimir potential.
\ As in the electrostatic case, we employ a parallel plate approximation
leading to $U_{C}\left( s\right) =\left( -\pi ^{2}/720\right) \left( \hbar
cA/s^{3}\right) $, where $\hbar $ is the Planck's constant and $c$ is the
velocity of light \cite{850}. \ Using these expressions we rewrite $U_{t}$
in a dimensionless form:

\begin{equation}
U_{t}=U_{E}\left( g\right) \left[ \left( \frac{x}{g}\right) ^{2}-\frac{%
\alpha }{1-x/g}-\frac{\beta }{1-\left( x/g\right) ^{3}}\right] ,
\end{equation}
where $\alpha =-U_{V}\left( g\right) /U_{E}\left( g\right) =\epsilon
_{0}AV^{2}/kg^{3}$ represents the relative importance of the electrostatic
interaction (compared with the elastic term), and similarly $\beta
=-U_{C}\left( g\right) /U_{E}\left( g\right) =\left( \pi ^{2}/360\right)
\left( \hbar cA/kg^{5}\right) $ for the ratio of Casimir interaction to the
elastic term.

For relatively small values of $\alpha $ and $\beta $ (see figure 4(b)) the
potential energy has a local minimum at $x=x_{1}$ associated with a
metastable state. \ Figure 4(c) shows a typical example of $U_{t}$ in the
metastable regime. \ The angular frequency of small oscillations around the
local minimum is $\omega _{1}$. \ The barrier that separates the local
minimum\ at $x=x_{1}$ from the global minimum at $x=g$ has height $U_{b}$. \
Below, we calculate the parameters $\omega _{1}$ and $U_{b}$\ for two special
cases relevant to the present experiment.

Consider first the case of $\alpha \ll 1$ and $\beta \ll 1$. \ In this case
we find $x_{1}/g=\left( \alpha +3\beta \right) /2$ and $\omega _{1}/\omega
_{0}=\left( 1-\alpha /2-3\beta \right) $. \ Since the escape rate for this
case is very small, we do not estimate the associated $U_{b}$.

In the next case we consider the instability occurring with a strong
electrostatic and a weak Casimir interaction. \ Note first that if $\beta =0$
then a local minimum and a potential barrier exist only for $\alpha <8/27$.
\ At the critical point, $\alpha =8/27$, the local minimum and the local
maximum points coincide at $x/g=1/3$. \ We therefore expand the potential
near that point, assuming $\epsilon \equiv 8/27-\alpha \ll 1$ ($\epsilon >0$%
) and also $\beta \ll 1$. \ To lowest order we find

\begin{mathletters}
\begin{eqnarray}
x_{1}/g &=&1/3-\left( 1/8\right) \sqrt{32\epsilon -216\beta }  \label{es_ins}
\\
\omega _{1}/\omega _{0} &=&\left( 3/4\right) \left( 32\epsilon -216\beta
\right) ^{1/4} \\
U_{b}/kg^{2} &=&\left( 3/512\right) \left( 32\epsilon -216\beta \right)
^{3/2}
\end{eqnarray}
The critical point is thus shifted to $\alpha =8/27-27\beta /4$ due to the
Casimir effect. \ The nonlinear frequency-shift parameter $\kappa $ is
calculated from the coefficients of the Taylor expansion of the potential
around the point $x=x_{1}\cite{mech}$. \ To lowest order we find $\kappa
g^{2}\propto \left( 32\epsilon -216\beta \right) ^{-3/4}$.

To compare the above results with theory we substitute the area, $A,$ with
an effective area, $A_{eff}=2.3A$, chosen to ensure that the measured and
calculated critical points coincide. \ This substitution serves to account,
in part, for the crude approximations employed in the model. \ The solid
lines in Fig. 2(a) and 2(b) show the theoretical results of Eq. (\ref{es_ins})%
. \ Agreement with experiment for both $x_{1}$ and $\omega _{1}$ is
moderately good despite the simplicity of the model. \ Note that our
measurements within the nonlinear regime are strongly affected by
nonlinearity in the detector's response. \ Given that this is not properly
accounted for in our model, we do not attempt comparison with theory for the
parameter $\kappa $.

We now turn our attention to the lifetime of the metastable mechanical state
before pull-in. \ At finite temperatures, thermal excitation above the
energy barrier can allow the system to ''escape'' from the freely-moving to
the pulled-in state. \ The rate for this generic type of process was first
found by Kramers\cite{284}, via the Fokker-Planck equation. \ The thermal
escape rate for the case of small damping is

\end{mathletters}
\begin{mathletters}
\begin{equation}
\Gamma _{t}=\frac{\omega _{1}}{Q}\frac{U_{b}}{k_{B}T}\exp \left( -\frac{U_{b}%
}{k_{B}T}\right) .  \eqnum{6}  \label{thermal}
\end{equation}
This result is valid when $U_{b}/k_{B}T\gg 1$ and $U_{b}/k_{B}T\gg 1/Q$. \
Alternatively, at low temperatures the system escapes via quantum tunneling 
\cite{cle}. \ Caldeira and Leggett (CL) \cite{211} \cite{374} have explored
what is termed {\it macroscopic quantum tunneling (}MQT{\it )} in systems
with many degrees of freedom. In the case of small damping they find that
the MQT rate is

\end{mathletters}
\begin{mathletters}
\begin{equation}
\Gamma _{Q}=\omega _{1}\xi _{1}\sqrt{\frac{U_{b}}{\hbar \omega _{1}}}\exp %
\left[ -\frac{U_{b}}{\hbar \omega _{1}}\left( \xi _{2}+\xi _{3}/Q\right) %
\right] .  \eqnum{7}  \label{quantum}
\end{equation}
Here $\xi _{1}$, $\xi _{2}$ and $\xi _{3}$ are positive constants of order unity. \
The rate for the case of no dissipation ($1/Q=0$) coincides with the
conventional WKB\ result for tunneling within a noninteracting,
single-particle picture. \ The CL\ model yields a description of the effect
of dissipation upon MQT involving only a single parameter, $Q$, that
subsumes all of the precise details of the interaction between the beam and
the environment. \ Equation (\ref{quantum}) explicitly demonstrates that the
effect of dissipation is to suppress the tunneling rate.

In the present experiment the temperature is relatively high ($k_{B}T\gg
\hbar \omega _{1}$) therefore the rate of thermal escape is much larger
compared to the rate of quantum tunneling. \ However, the opposite
situation, namely $\Gamma _{Q}\gtrsim \Gamma _{t}$, may be realized with
present-day technology. \ For high frequency mechanical oscillators (see for
example \cite{224}) the quantum limit, $k_{B}T\lesssim \hbar \omega _{1}$, 
is reachable at cryogenic temperatures. \ The model employed in the present
paper may be useful in designing an experiment to study phenomena in this
regime.

Comparison between the simple theory and experiment indicates significant
disagreement. \ We extract the parameter $32\epsilon -216\beta $ from Eq.
(5b) and the value $\omega _{1}/\omega _{0}=0.81$ found experimentally when
the pull-in occurs. \ This parameter is substituted in Eq. (5c) to find $%
U_{b}$. \ This yields $U_{b}/k_{B}T\simeq 10^{7}$, a value which, when
substitited into Eq. (\ref{thermal}), yields a lifetime that is essentially
infinite for all practical purposes. \ Experimentally, however, we observed
pull-in after only a few minutes, in striking contradiction with this
theoretical estimate. \ Similar devices, having $g=$1.29, 0.71 and 0.65 $\mu 
$m, exhibited pull-in when $\omega _{1}/\omega _{0}=$ 0.84, 0.987, and 0.995
respectively. In these cases, again, the simple theory is far from
explaining what is experimentally observed. \ We note, in general, that any
theory predicting $U_{b}=Akg^{2}\left( \omega _{1}/\omega _{0}\right) ^{B}$
near the critical point, where both $A$ and $B$ are of order unity, will
exhibit similar discrepancy. \ This is due to the exponential dependence in
Eq. (\ref{thermal}), in concert with the fact that $kg^{2}/k_{B}T\simeq
10^{7}\gg 1$. \ 

Our results do not appear to be anomalous. \ We have analyzed data from
previous experiments involving CTR devices made from stress-free silicon\cite
{15}, in which pull-in was also observed upon application of a DC voltage
between electrodes. \ Calculating the expected lifetime for the experimental
conditions of Ref. \cite{15} using Eq. (\ref{thermal}), we find lifetimes
that are in similarly gross contradiction with experimental observations. \
In principle, slow processes of stress relaxation in both cases may be the
underlying mechanism inducing early pull-in \cite{3415}. \ Further
investigations are warranted. \ In our work we have taken care to rule out
spurious effects from environmental electrical noise sources that could, in
principle, drive the system out of the metastable state. \ Measured
electrical noise was far smaller than the steady bias voltages applied. \
Also, ''early'' pull-in occurred even when the electron beam was blanked,
ruling out the possibility that some form of parasitic local charging may be
operative. \ 

The present work shows that MEMS can provide an ideal approach to studies of
mechanical metastability. \ Further experimental and theoretical work will
elucidate the underlying mechanisms governing the instability of these
systems. \ Such work is important not only from a theoretical point of view,
but also from the standpoint of practical device engineering. \ Moreover,
further enhancement of experimental techniques will enable new studies of
macroscopic quantum tunneling in mechanical systems.

The authors are grateful to K. Schwab for his assistance in sample
fabrication. \ This research was supported by DARPA MTO/MEMS under grant
DABT63-98-1-0012. \ E.B. gratefully acknowledges suppport from a Rothschild
Fellowship and the R.A. Millikan Postdoctoral Fellowship at Caltech.

\bigskip

\newpage 
\begin{figure}[tbp]
\caption{(a) Side view micrograph of the device. \ (b) External force is
applied by a voltage bias and motion is detected by focusing the e-beam on a
point on the edge of the structure and measuring the secondary reflected
electrons.}
\label{fig:fig1}
\end{figure}

\begin{figure}[tbp]
\caption{(a) Displacement of the center of the beam $x_{1}$ divided by $g$.
\ (b) Frequency shift $\Delta \protect\omega $ divided by the unperturbed
frequency $\protect\omega _{0}$. \ (c) \ Nonlinear parameter $\protect\kappa %
g^{2} $.}
\label{fig:fig2}
\end{figure}

\begin{figure}[tbp]
\caption{The mechanical response as a function of frequency around the
fundamental resonance with $V_{dc}=20$ V and $V_{ac}=$25, 50, 75,...,225 mV.
\ The inset shows the hysteresis loop for $V_{ac}=225$ mV when the frequency
is scanned forth and back across the resonance.}
\label{fig:fig3}
\end{figure}

\begin{figure}[tbp]
\caption{(a) A one dimensional model describing our system. \ (b) the
metastable regime in the $\protect\alpha \protect\beta $ plane. \ (c) The
potential energy in the metastable regime has a barrier $U_{b}$ separating
the metastable state at $x=x_{1}$ and the contact state at $x=g$.}
\label{fig:fig4}
\end{figure}
\end{mathletters}

\end{document}